# Weber's law implies neural discharge more regular than a Poisson process

Jing Kang,[1] Jianhua Wu,[2] Anteo Smerieri[3] and Jianfeng Feng[1]
[1]Department of Computer Science, University of Warwick, Coventry CV4 7AL, UK
[2]Department of Neuroscience, Columbia University, New York, NY 10032, USA
[3]Department of Physics, University of Parma, Viale Usberti 7A, Parma 43100, Italy

*Keywords*: attractor network, interspike interval, psychophysical law, spike rate

## Abstract

Weber's law is one of the basic laws in psychophysics, but the link between this psychophysical behavior and the neuronal response has not yet been established. In this paper, we carried out an analysis on the spike train statistics when Weber's law holds, and found that the efferent spike train of a single neuron is less variable than a Poisson process. For population neurons, Weber's law is satisfied only when the population size is small (< 10 neurons). However, if the population neurons share a weak correlation in their discharges and individual neuronal spike train is more regular than a Poisson process, Weber's law is true without any restriction on the population size. Biased competition attractor network also demonstrates that the coefficient of variation of interspike interval in the winning pool should be less than one for the validity of Weber's law. Our work links Weber's law with neural firing property quantitatively, shedding light on the relation between psychophysical behavior and neuronal responses.

## Introduction

It is of little doubt that there exists a relation between the exquisite psychophysical sensitivity of human and animal observers and the sensitivity of individual cortical neurons. The transformation between sensory cortical neurons signals and the perceptual responses remains unclear, despite the fact that the link between the neuronal activity and psychophysical judgment of sensory processing has been intensively studied by many researchers (Shadlen & Newsome, 1994; Sawamura *et al.*, 2002). The idea of quantitatively relating cortical neuronal activities to sensory experiences was first proposed by Werner & Mountcastle (1963), who enunciated some fundamental principles for the analysis of neuronal discharge in a psychophysical context. Weber's law (also called Weber–Fechner law; Fechner *et al.*, 1966), one of the classical psychophysical laws, states that the ratio between the just noticeable differences (JNDs) in stimulus intensity ($\Delta I$) and the reference stimulus intensity ($I$) is a constant $k$ (Weber's constant), i.e. $\Delta I/I = k$. This phenomenon has been observed in a wide range of moderately intense stimuli experiments in sensory perception in terms of weights (Fechner *et al.*, 1966), pure tones (Gescheider *et al.*, 1990), light intensities (Wald, 1945), sizes (Smeets & Brenner, 2008), texture roughness (Johnson *et al.*, 2002), numbers (Dehaene, 2003), etc., but a link between this psychophysical property and neuronal activity is still lacking.

Weber's law describes the relationship only between the stimulus intensity and psychophysical behavior, so the challenge to study this law at the neuronal level is how to characterize the unclear intermediate connections of stimulus–neuronal and neuronal–psychophysical responses. In most biophysical and psychophysical experiments, the relation of neural response rate and input stimulus intensity generally follows a non-linear sigmoid function. The middle range of a sigmoid function is asymptotically a straight line reflecting the linear relation between neural firing and the stimulus intensity. Starting from the analyses on the simplest linear case of the input–output relation between the stimulus intensity and neuronal response rate, we further extend our analysis on the non-linear input–output relation (sigmoid function). Under Weber's law, it is found that for both linear and non-linear relations of input stimulus and output neuronal responses, the final results are similar in terms of the neuronal spiking process. For a more biological realistic setup on neuronal input–output relation, we also investigate the neuronal spike train properties in the spiking network model when Weber's law holds. Therefore, we can establish the intermediate link between the psychophysical law (Weber's law) and neuronal spike train statistics.

In this paper, we theoretically derived a relationship between the mean ($\mu$) and the standard deviation ($\sigma$) of the neuronal spike rate when Weber's law holds, and express their relation in terms of the dispersion of interspike intervals (ISIs) that require $CV_{ISI} \in [0.5, 1]$. Starting from single neurons, we studied the independent and correlated superimposed population neuronal discharge patterns, as well as competition attractor network neurons. The competitive attractor neural network also indicates that the neuronal ISI should be more regular than a Poisson process in the winning pool so that Weber's law holds. Our work links Weber's law with neural firing property quantitatively: Weber's law indicates the variability of the neuronal spike train; meanwhile, given a series of spike train data

*Correspondence*: Professor J. Feng, as above.
E-mail: Jianfeng.Feng@warwick.ac.uk

stimulated at different intensities, we can determine whether this psychophysical law is satisfied.

## Materials and methods

### First and second order statistics under Weber's law

Applying a constant stimulus $I$ to a single neuron repeatedly, the neuron will fire at a mean rate $\mu$ with variance $\sigma^2$ over a certain stimulus time. If the increment of input stimulus intensity $\Delta I$ is just noticeable, the mean output firing rates $\mu$ and $\mu + \Delta\mu$ should be statistically discriminable under some criterion $\varepsilon$. We firstly assume the linear relation between the stimulus intensity $I$ and the output neuronal mean discharge rate $\mu$ (spikes/s), as linearity of the input–output relation between stimulus intensity and neuronal response rate is widely accepted and intensively used in simulation modeling (Holt & Koch, 1997), and also supported by experiments (Leng *et al.*, 2001; Enoki *et al.*, 2002; Johnson *et al.*, 2002).

Therefore,

$$\mu = aI, \quad \text{where a is the scale,}$$

and

$$\frac{\Delta\mu}{\mu} = \frac{a\Delta I}{aI} = \frac{\Delta I}{I} = k. \tag{1}$$

where $k$ is Weber's constant.

Define two distributions with different means and variances ($\mu_1$, $\sigma_1{}^2$) and ($\mu_2$, $\sigma_2{}^2$), respectively. Without loss of generality, assume that $\mu_1 < \mu_2$, and that $\mu_1$ and $\sigma_1$ follow the same relation as $\mu_2$ and $\sigma_2$. In the Supporting information, Appendix S1, it is shown that

$$x_0 = \frac{\sigma_2}{\sigma_1 + \sigma_2}\mu_1 + \frac{\sigma_1}{\sigma_1 + \sigma_2}\mu_2$$

can be the discriminant that minimizes the misclassification rate $\varepsilon$ for any two distributions with known ($\mu_1$, $\sigma_1{}^2$) and ($\mu_2$, $\sigma_2{}^2$). {Discriminant is a criterion for classifying an 'individual' into one of two or more populations based on observations on that individual. In this case, if an observed data $x$ is less than the discriminant $x_0$, $x$ is classified as group 1 [with distribution ($\mu_1$, $\sigma^2{}_1$)], and if $x$ is greater than $x_0$, $x$ is classified as group 2 [with distribution ($\mu_2$, $\sigma_2{}^2$)]}. To begin with a simple situation, assume that the neuronal spiking rate of a single neuron follows normal distribution with parameters $\mu$ and $\sigma^2$. Thus, the misclassification rate is

$$\varepsilon = \frac{1}{2}\left(1 + \mathrm{erf}\left(\frac{x_0 - \mu_2}{\sqrt{2}\sigma_2}\right)\right) + \frac{1}{2}\left(1 - \mathrm{erf}\left(\frac{x_0 - \mu_1}{\sqrt{2}\sigma_1}\right)\right)$$

for two normal distributions N($\mu_1$, $\sigma_1{}^2$) and N($\mu_2$, $\sigma_2{}^2$), where the error function erf(.) is defined as

$$\mathrm{erf}(x) = \frac{2}{\pi}\int_0^x e^{-t^2}dt.$$

Therefore, we have

$$\mathrm{erf}(x_2) - \mathrm{erf}(x_1) = 2(\varepsilon - 1),$$

where

$$x_1 = \frac{x_0 - \mu_1}{\sqrt{2}\sigma_1} \quad \text{and} \quad x_2 = \frac{x_0 - \mu_2}{\sqrt{2}\sigma_2}.$$

By substituting $x_0$ into $x_1$ and $x_2$, we have $x_1 = -x_2$, and because the error function is an odd function, we have

$$\mathrm{erf}(x_2) = \varepsilon - 1.$$

One of the good approximation forms for the error function is

$$\mathrm{erf}(x) \approx \sqrt{1 - \exp\left(-\left(\frac{2x}{\sqrt{\pi}}\right)^2\right)},$$

and the detailed derivation of this approximation is presented in supporting Appendix S1. After a simple calculation, the relation between $\mu_1$, $\mu_2$, $\sigma_1$ and $\sigma_2$ becomes

$$\mu_2 - \mu_1 = C(\sigma_1 + \sigma_2), \tag{2}$$

where

$$C = \sqrt{\frac{\pi}{2}\mathrm{In}\left(\frac{1}{1 - (\varepsilon - 1)^2}\right)} \tag{3}$$

is a constant determined by the misclassification rate $\varepsilon$.

To derive the relation between $\sigma$ and $\mu$ for normal distribution, assume $\sigma$ is a function of $\mu$, i.e. $\sigma = f(\mu)$. Substitute $\sigma_1 = f(\mu_1)$ and $\sigma_2 = f(\mu_2)$ into Eq. (2), and apply the first-order Taylor expansion on function $f$ at point $\mu$, we have

$$C(2f(\mu) + f'(\mu)(\mu_1 + \mu_2 - 2\mu)) = \mu_2 - \mu_1. \tag{4}$$

From Weber's law, the relation for $\mu_1$ and $\mu_2$ obeys $(\mu_2 - \mu_1)/\mu_1 = k$. Substitute $\mu_2 = (k + 1)\mu_1$ in Eq. (4) and let $\mu_1 = \mu$, it yields

$$f'(\mu) + \frac{2}{k\mu}f(\mu) - \frac{1}{C} = 0. \tag{5}$$

This is a first-order non-homogeneous ordinary differential equation (ODE), which has a general solution

$$\sigma = f(\mu) = \frac{k}{C(k+2)}\mu + \mu^{-\frac{2}{k}}c_0, \tag{6}$$

where $c_0$ is a constant determined by the initial condition.

When the neural response rate has a non-linear relation with respect to the input stimulus intensity (e.g. sigmoid function), the analysis is relatively complicated but the theoretical solution can still be obtained (see Discussion for details).

### Superposition process

The superposition process (or superimposed process) $N(t)$ is defined as the total number of arrivals for all neurons that occur up to time $t$:

$$N(t) = \sum_{i=1}^{p} N_{t,i} \quad t \geq 0,$$

where $N_{t,i}$ is the spike count for the $i$th neuron during the time interval [0, $t$]. Assume that each neuron in the population is identical and independently following Gamma distribution $\Gamma(A, B)$. Even though the expression of the density function of the superimposed ISI is complicated (Cox & Miller, 1965; Lawrance, 1973), the superimposed counting rate statistics (mean and variance) in a small time window $W$ can be found theoretically.

Define the correlation among network neurons by the spike train correlation between pair-wised cells spike counts $n_i$ and $n_j$ over a sliding window of length $W$:

$$\rho_W = \frac{\mathrm{cov}(n_i, n_j)}{\sqrt{\mathrm{var}(n_i)\mathrm{var}(n_j)}}.$$

The superposition process of correlated spike train of population neurons of size $p$ over sliding window $W$ can be found as:

$$Y = \mu + \frac{1}{p}\mathbb{1}_{1\times p} M^{\frac{1}{2}}(\mathbf{X} - \boldsymbol{\mu}),$$

where $Y$ stands for the standard superimposed spike count and

$$\mathbf{X} = \begin{pmatrix} x_1 \\ \vdots \\ x_p \end{pmatrix}$$

is the spike rate of each neuron in the network with

$$E(\mathbf{X}) = \mu\mathbb{1}_{p\times 1} = \boldsymbol{\mu} \text{ and}$$

$$\mathrm{cov}(\mathbf{X}) = \sigma^2 \mathbf{I}_{p\times p,}$$

($\mathbb{1}_{p\times 1}$ is the $p$-dimensional column vector with each element as 1 and $I_{p\times p}$ is the $p$-dimensional identical matrix) and $M$ is the correlation matrix of the form

$$M = \begin{bmatrix} 1 & \rho & \cdots & \rho \\ \rho & 1 & \rho & \vdots \\ \vdots & \rho & \ddots & \rho \\ \rho & \cdots & \rho & 1 \end{bmatrix}.$$

Therefore, the standard superimposed spike count rate is

$$\mathrm{E}(Y) = \mu, \text{and}$$

$$\mathrm{var}(Y) = \frac{\sigma^2}{p}(1 + (p-1)\rho).$$

If $\rho = 0$, this is an independent superposition process $Y$ with $E(Y) = \mu$, and $\mathrm{var}(Y) = \sigma^2/p$.

### *Competition attractor network*

The model of a competition-based network for decision making was originally proposed by Brunel & Wang (2001), and further studied by Deco & Rolls (2006). The task of the network is to make a decision between two possible alternatives, according to the characteristics of a sensory input, by reaching one of two predetermined firing states. A typical task is the comparison of two different stimuli, for example vibrotactile stimulation frequency.

The network is composed of four pools of fully connected leaky integrate-and-fire neurons, both excitatory and inhibitory. The pools are divided according to the strength of the connections between the neurons (Fig. 7A). Each pool receives external inputs in the form of excitatory spikes with a Poisson distribution; the frequency of the inputs depends on the stimuli characteristic to be compared. A decision is reached when one of two specialized excitatory neurons pools (pool A or pool B) reaches a high-frequency (30–60 Hz) firing state, while the other is almost silent. Competition is made possible by a pool of inhibitory neurons, which usually fire at about 20 Hz. The inhibitory pool suppresses the activity of one of the two specialized pools, while the non-specific pool consists of non-specialized excitatory neurons that do not react to the stimuli characteristics. More details on the network architecture can be found in supporting Appendix S1 and Wang (2002).

The network reaches a correct decision when the high-rate firing pool is the one with the larger input frequency; otherwise the decision is considered ‘wrong’. Deco & Rolls (2006) have shown that for a certain input range the network follows Weber’s law, in the sense that the difference between input frequencies required to achieve, over many trials, a certain success rate (85% in this paper) is proportional to the amplitude of one of the two input frequencies.

We rebuilt the competition-based neural network model (Fig. 7A) and measured the value of $\mathrm{CV_{ISI}}$ for each pool, verifying that Weber’s law holds for our implementation of the model. The input ($F_{\mathrm{in}}$) to one of the specialized pools (e.g. pool A) is considered as the reference input (Fig. 7A), while the reference input frequency $\Delta F_{\mathrm{in}}$ was chosen in a range that allows the network not to be saturated by the inputs. The input frequency $F_{\mathrm{in}} - \Delta F_{\mathrm{in}}$ applied to pool B is set between 30% and 100% of the reference input frequency (thus, $\Delta F_{\mathrm{in}}$ varied between 0 and $0.7{*}F_{\mathrm{in}}$). For each pair of input frequencies, 200 simulations were run, and the success rate achieved for each pair was recorded. A curve of the success rate vs $\Delta F_{\mathrm{in}}$ can be drawn for each reference value $F_{\mathrm{in}}$. By fitting the curve for each value of $F_{\mathrm{in}}$, the $\Delta F_{\mathrm{in}}$ value achieving a certain success rate can be found.

We then altered the input spike train so that Weber’s law does not hold for the discrimination task, by applying a different input spike distribution. The distribution of the interspike input intervals was altered to be uniform between 0 and twice the average interval, which is the reciprocal of the input frequency. The rest of the network setup remains the same.

## Results

### *Weber’s law in Gaussian-distributed firing rate*

In the Materials and methods, we firstly derived the relation between the mean ($\mu$) and standard deviation [$\sigma = f(\mu)$] of the neuronal discharge rate from normal distribution when Weber’s law holds in terms of first-order non-homogeneous ODE [Eq. (5)], with the Weber’s constant $k$ ranging from 0.05 to 0.3 (Gescheider *et al.*, 1990). The constant $C$ [Eq. (3)] is determined by the misclassification criterion ε, and the value of ε (ranging from 5 to 20%) is not crucial to the final result. ε is fixed to be 15% (Deco & Rolls, 2006) in this paper, and as a result $C \approx 1.4$. In the solution [Eq. (6)] of the first-order non-homogeneous ODE, the second term $\mu^{-2/k}c_0$ can be neglected (as $k$ is much smaller than 1, and the mean discharge rate $\mu$ is fixed within 1–200 spikes/s), so the standard deviation $\sigma$ and the mean $\mu$ of the discharge rate have a linear relation

$$\sigma = \frac{k}{C(k+2)}\mu. \tag{7}$$

We call Eq. (7) Weber’s equation. On the one hand, Weber’s equation is derived from Weber’s law; on the other hand, Weber’s law is satisfied when the mean and standard deviation of the neuronal discharge rate obey Weber's equation. More detailed theoretical derivations can be found in the Materials and methods, and supporting Appendix S1. Moreover, even though this linear relation between $\sigma$ and $\mu$ in Weber's equation is derived from normal distribution, this result can be generalized to any distribution, even a distribution with only known mean and variance, by just varying the expression of the scaling parameter $C$. The detailed generalization of this linear relationship for any distribution is presented in the supporting Appendix S1.

A simulation for this normally distributed spiking rate with its mean $\mu$ and STD $\sigma$ satisfying Weber's equation is shown in Fig. 1A for different values of $k$. The simulated ratio of $\Delta\mu/\mu$ is smaller than the given Weber constant $k$ (simulated slopes = 0.04, 0.07, 0.14 and 0.22, respectively). This smaller ratio can be caused by the truncation error at the higher order of the Taylor expansion and the approximation of the error function (see Materials and methods). However, Weber's law provides a perfect description of linearity between the JNDs $\Delta\mu$ and reference firing intensity $\mu$.

We have also demonstrated the relation between JNDs $\Delta\mu$ and $\mu$ when they do not follow Weber's equation as counterexamples. In Fig. 1B, when the standard deviation $\sigma$ of the normal distribution is constant, consequentially the JNDs $\Delta\mu$ is a constant value for any reference rate $\mu$; the reason for this is that the normal distribution with different means but constant variance is just a shift without changing its shape. Thus, for fixed misclassification rate ($\varepsilon$ = 15%), $\Delta\mu$ is always the same for discrimination. When the STD ($\sigma$) equals the square root of the mean $\mu$ ($\sigma = \mu^{1/2}$; Fig. 1C), and in turn $\Delta\mu$ has a non-linear relation with respect to $\mu$ ($\Delta\mu = 2C\mu/(\sqrt{\mu} - C/2)$, the plausible range of the scale parameter in Weber's equation ($a = b\mu$) is $b \in [0.02, 0.1]$[as $C \approx 1.4$ in Eq.(7)]. If coefficient parameter $b$ goes beyond the plausible range ($\sigma = \mu/2$; Fig. 1D), the JNDs $\Delta\mu$ is still a linear function of the reference rate $\mu$, but with the slope $k' \approx 2.1$ (as the slope $k' = 2bc/(1 - bc)$, $b = ½$ in Fig. 1D), which is much larger than realistic psychophysical values ($k \in [0.05, 0.3]$).

### *Weber's law in single neuronal ISI*

Many researchers have pointed out that Poisson or renewal process is more appropriate to describe the neuronal firing activity (Cox & Smith, 1954; Cox & Miller, 1965). Assume that the interspike interval (ISI) $T$ of a spiking neuron follows Gamma distribution, i.e. $T \sim \Gamma(A, B)$, with mean $E(T) = AB$ and variance $\mathrm{var}(T) = AB^2$. Exponential distribution is a special case of Gamma distribution when the parameter $A = 1$. The coefficient of variation of the ISI ($CV_{ISI}$) equals $CV_{ISI} = \mathrm{STD}(T)/E(T)$.

If ISIs follow Gamma distribution, the corresponding neural spike rate can be described by a renewal process, and the spike rate has a mean

$$\mu = E\left(\frac{N(t)}{t}\right) = \frac{t}{tE(T)} = \frac{1}{E(T)}$$

and variance

$$\sigma^2 = \mathrm{var}\left(\frac{N(t)}{t}\right) = \frac{\mathrm{var}(T)t}{E^3(T)}\frac{1}{t^2} = \frac{\mathrm{var}(T)}{E^3(T)t} = CV_{ISI}^2\frac{1}{E(T)t}$$

(Cox & Isham, 1980). Expressing Weber's equation by $CV_{ISI}$, we obtain

$$CV_{ISI} = \sigma\sqrt{E(T)t} = \sigma\sqrt{\frac{t}{\mu}} = \frac{k}{C(k+2)}\mu\sqrt{\frac{t}{\mu}} = \frac{k}{C(k+2)}\sqrt{\mu t}.$$

The above expression describes the linear relation between $CV_{ISI}$ and the square root of spike count over time period $t$. As we usually use Hertz ($t$ = 1 s) as the unit to quantify neuronal spiking, we can write

$$CV_{ISI} = \frac{k}{C(k+2)}\sqrt{\mu}. \qquad (8)$$

If $CV_{ISI} = 1$, the neural discharge follows the Poisson process. If $CV_{ISI} > 1$, we call the renewal process the 'super-Poisson' process;

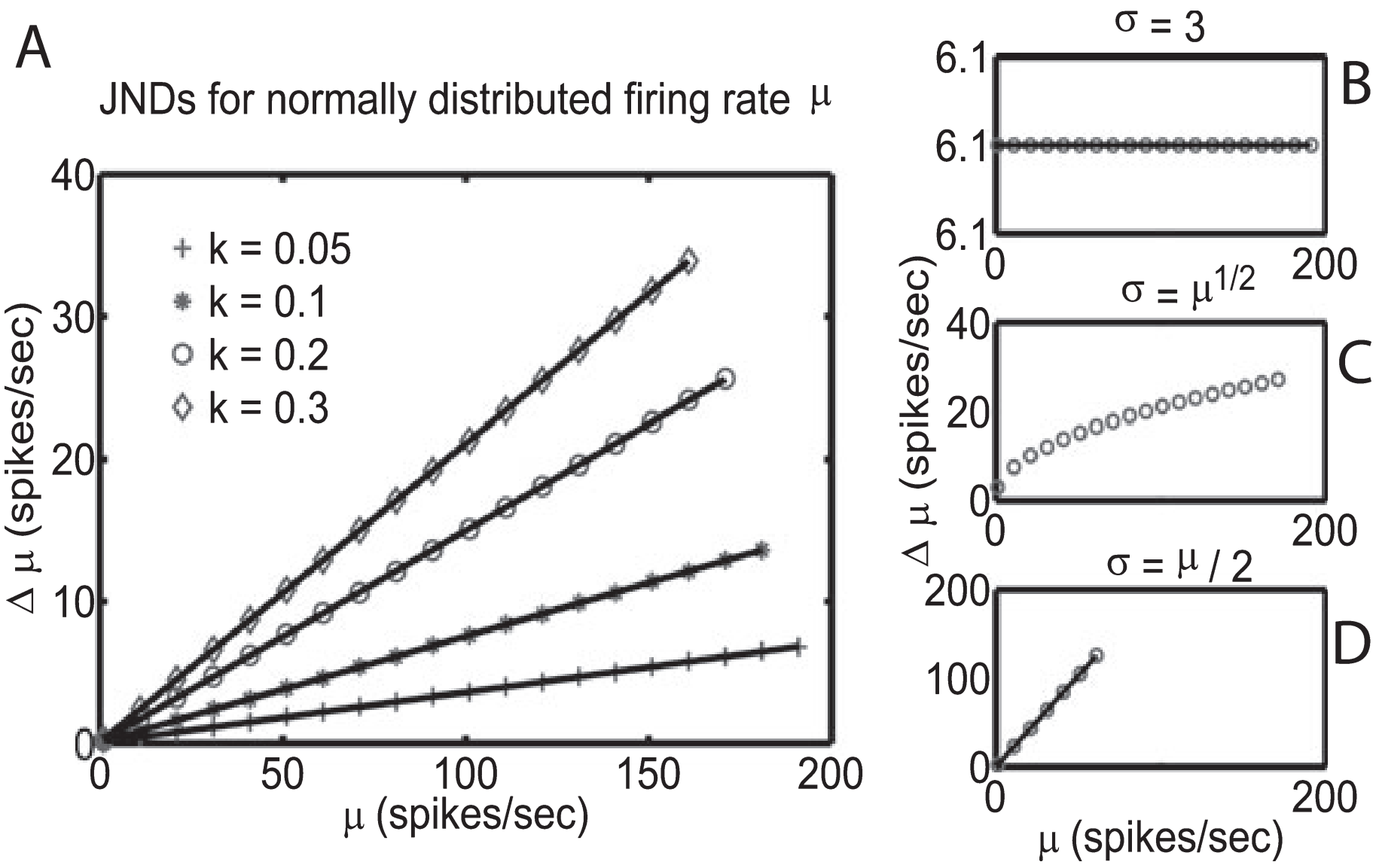

FIG. 1. The relation of just noticeable differences (JNDs) and reference spiking intensity $\mu$ for normally distributed firing rate under different $\sigma$–$\mu$ relations. (A) Under Weber's equation, the JND is a linearly increasing function of the firing rate with different slopes at different values of Weber's constant $k$. The simulated JNDs have a relatively smaller slope than $k$, possibly raised from the truncation error from the approximation form. (B–D) Examples for $\sigma$–$\mu$ relation not satisfying Weber's equation, where (B) STD is constant $\sigma$ = 3, (C) $\sigma = \mu^{½}$ and (D) $\sigma = \mu/2$.

when $CV_{ISI} < 1$, it is the 'sub-Poisson' process [in physics, the super-Poisson process is defined by the index of dispersion variance-mean ratio (VMR) of the counting of events with VMR > 1, and sub-Poisson is similarly defined as VMR < 1 (Kolobov, 1999) besides, $VMR = CV^2_{ISI}$ in the renewal process]. For a very regular spike train ('pacemaker'), the ISI histogram has a very narrow peak and $CV_{ISI} \rightarrow 0$. In the case of a random spike train (Poisson process), the ISI are exponentially distributed and $CV_{ISI} = 1$. The $CV_{ISI}$ can be larger than one in the case of a multistate neuron (Wilbur & Rinzel, 1983).

One may argue that the positive correlation between the firing rate $\mu$ and $CV_{ISI}$ contradicts the experimental observation because of the existence of the refractory period of each neuron, and in real neurons the $CV_{ISI}$ should decrease as the neuronal firing rate increases. However, the real neurons usually fire at the rate between 0 and 40 Hz, seldom up to 100 Hz, so it is no surprise to see that the variability of spike train increases as the neuronal firing rate increases under certain circumstances. Besides, the existence of the refractory period does not influence our theoretical results at all, because the neuronal spiking process is still regarded as a renewal process. (Please refer to the supporting Appendix S1 for detailed analysis.)

The range of $CV_{ISI}$ under Weber's law can be determined from the range of Weber's constant ($k \in [0.05, 0.3]$)

$$\frac{\sqrt{\mu}}{100} < CV_{ISI} = \frac{k}{C(k+2)}\sqrt{\mu} < \frac{\sqrt{\mu}}{10}, \tag{9}$$

where $C = 1.4$. Define Inequality (9) as 'Weber's range' for a single neuron. The detailed relations between the parameters are presented in Table 1. For a single neuron, $CV_{ISI}$ is rather small (0.14–0.3) for biological feasible firing rate ($10 \le \mu \le 200$ Hz), which means the neuron fires very regularly. However, the small value of $CV_{ISI}$ contradicts the irregularity of the cortical neuronal discharge behavior *in vivo*.

We presented several simulation examples to demonstrate the range of the feasible firing rate under which Weber's law holds, by examining the spike count over a small sliding time window ($W = 20$ ms). In the simulation, the ISIs follow Gamma distribution with fixed parameter $A = 1/(CV_{ISI})^2$ and rate-dependent parameter $B = 1/(\mu A)$. Figure 2 demonstrates the relation between the JNDs of firing rate $\Delta\mu$ and $\mu$ for sub-Poisson, Poisson and super-Poisson discharge processes. For the sub-Poisson process (Fig. 2A), a linear relationship between $\Delta\mu$ and $\mu$ is more notable for a relatively small value of $CV_{ISI}$ (0.2) than that of large values of $CV_{ISI}$ (0.5 and 0.8), but the slope of the linear regression curve for $CV_{ISI} = 0.2$ is much smaller (0.02) than Weber's constant. For the Poisson process, Weber's law does not hold when the firing rate $\mu < 100$ Hz (Fig. 2B; Table 1), as the increasing trend of $\Delta\mu$ is non-linear with respect to $\mu$. In contrast, when $\mu > 100$ Hz, the relation between JNDs $\Delta\mu$ and reference firing rate $\mu$ is linear (slope = 0.06). For the super-Poisson process ($CV_{ISI} = 1.2$, 1.5 and 3; Fig. 2C), Weber's law cannot be satisfied because of the non-linearity of the curves. Moreover, the $y$-axis intercept of these curves does not pass the origin (0, 0). Figure 2D is the contour plot of Weber's constant $k$ vs. $CV_{ISI}$ and firing rate $\mu$. It illustrates the range of $CV_{ISI}$ and firing rate $\mu$ when $k$ lies within [0.05, 0.3].

Figure 3 shows the ISI distribution when $CV_{ISI}$ and firing rate $\mu$ follows Weber's equation. The ISI distribution bends towards longer ISI time for larger $CV_{ISI}$ because of the non-linear relation between $CV_{ISI}$ and the mean firing rate $\mu$ under Weber's equation [Eq.(7)].

From the experimental observations on cortical neuronal discharge variability, the $CV_{ISI}$ should be from 0.5 to 1, which is much bigger than the $CV_{ISI}$ value under Weber's range (0.14–0.3) for a single neuron. The discrimination behavior cannot be performed by single cortical neurons and, thus, population neurons need to be considered.

### *Superposition of independent neural discharge process*

We begin with the superimposed population neurons, with each neuron firing independently. Assume each individual neuron's firing pattern follows an identical distribution with mean discharge rate $\mu$ and standard deviation $\sigma$. If each neuron fires independently, the mean discharge rate $\mu$ after superposition is still the same as single neurons, but the variance of the instantaneous spike count becomes $\sigma^2/p$, where $p$ is the population size. Therefore, substitute $\mu$ and $\sigma^2/p$ into Weber's equation, we have

$$CV_{ISI} = \frac{k}{C(k+2)}\sqrt{\mu p}$$

TABLE 1. Parameter region under Weber's equation (with $0.05 \le k \le 0.3$)

| | Firing rate $\mu$ (Hz) | | | |
|---|---|---|---|---|
| Population size ($p$) | $CV_{ISI} = 0.5$ | $CV_{ISI} = 1$ | $CV_{ISI} = 1.5$ | $CV_{ISI}$ ($10 \le \mu \le 200$) |
| $p = 1$ | [25, 2500] | [100, $10^4$] | [225, 2.104] | [0.14, 0.3] |
| Correlation $\rho = 0$ | | | | |
| $p = 10$* | [3, 250] | [10, 1000] | [23, 2250] | [0.45, 1] |
| $p = 100$ | [0.25, 25] | [0, 100] | [2, 500] | [1.4, 3.2] |
| $p = 1000$ | [0.02, 3] | [0.1, 10] | [0.2, 23] | [4.5, 10] |
| Correlation $\rho = 0.1$ | | | | |
| $p = 10$ | [5, 472] | [19, 1900] | [42, 4200] | [0.3, 0.73] |
| $p = 100$* | [3, 270] | [11, 1100] | [25, 2500] | [0.43, 0.95] |
| $p = 1000$* | [3, 250] | [10, 1000] | [23, 2300] | [0.45, 1] |

This table describes the feasible range of the parameter values ($k$, $\rho$, $p$, $CV_{ISI}$ and $\mu$ under Weber's equation: $CV_{ISI} = k\sqrt{\mu p/1 + (p-1)p}/C(k+2)$. We mainly focus on the range of the firing rate $\mu$ at given $CV_{ISI}$ (0.5, 1 and 1.5) and the range of $CV_{ISI}$ at feasible firing rates ($10 \le \mu \le 200$), with the rest of the parameters fixed. It is concluded that the population size is $p \le 11$ for an independent superposition process and $p > 51$ for a correlated superposition process, under biologically feasible $CV_{ISI}$ ranging from 0.5 to 1, and that Weber's law is validated (see data rows population sizes marked with an asterisk'*'). ISI, interspike interval.

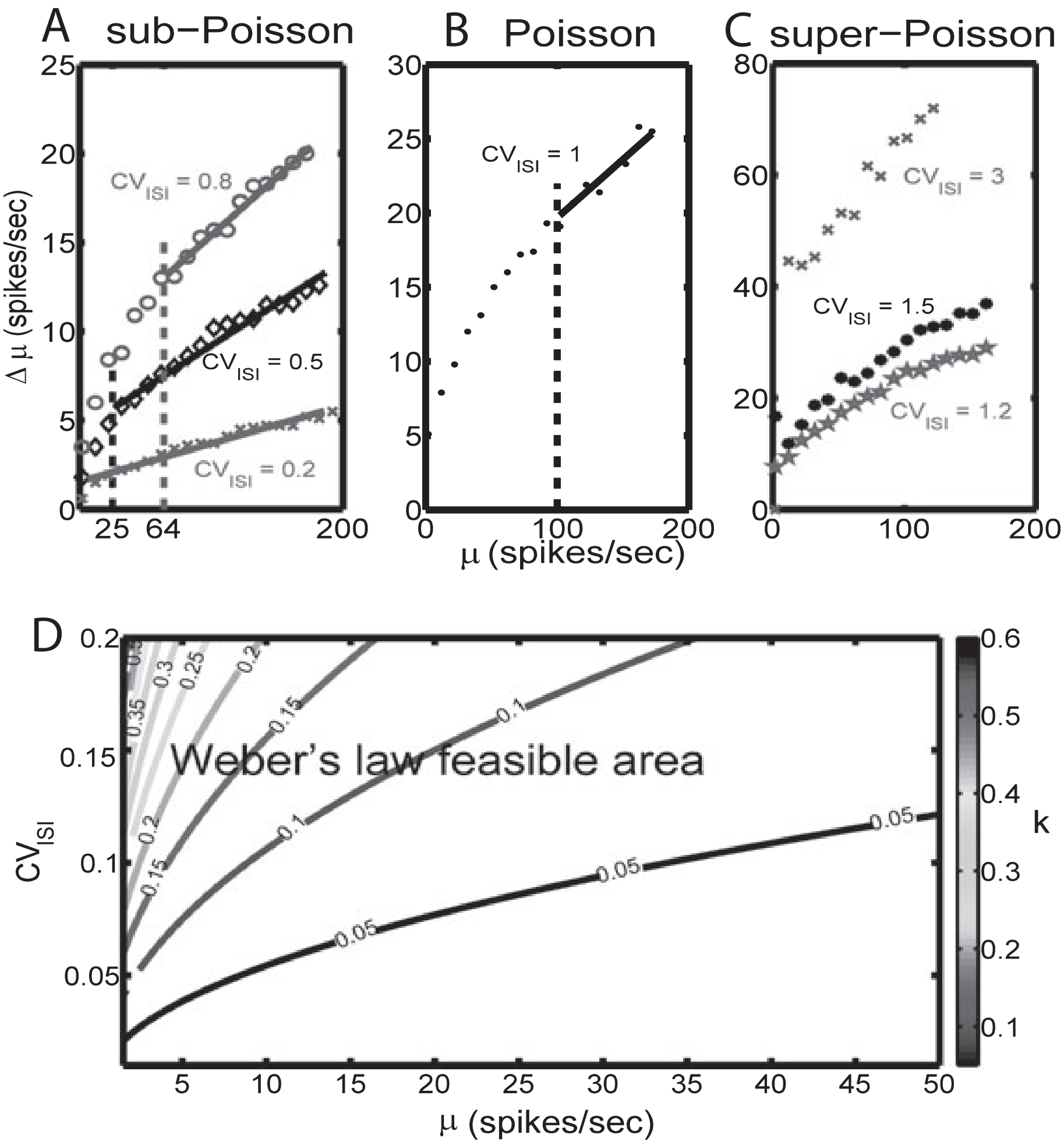

FIG. 2. The relation of JNDs and reference spiking intensity $\mu$ for (A) sub-Poisson, (B) Poisson and (C) super-Poisson neural discharge process. The vertical dashed line represents the lower bound of the feasible ranges of the neuronal firing rate under Weber's range. (D) The contour plot of Weber's constant $k$ indicating the feasible region of Weber's law in terms of the firing intensity ($\mu$) and dispersion of the interspike interval (ISI) ($CV_{ISI}$).

As a result, Weber's range can be derived from Inequality (9)

$$\frac{\sqrt{\mu p}}{100} < CV_{ISI} < \frac{\sqrt{\mu p}}{10},$$

where the value of $CV_{ISI}$ depends on neuronal discharge rate $\mu$ and population size $p$. The range for the firing rate $\mu$ can be determined for a given population size $p$ such that Weber's law holds (see Table 1 for detailed range of parameters).

The simulation results and illustrative parameter relation are shown in Figs 4 and 5A. The first row of Fig. 4 shows the case for a large population size ($p = 1000$). From Table 1, $CV_{ISI}$ has to be large if the population size is large ($CV_{ISI} > 4.5$). Thus, for a large population, each individual neuron should have a very irregular discharge process ($CV_{ISI} >> 1$) so that Weber's law is satisfied, but it is biologically unrealistic and impossible. From the simulation, if the value of $CV_{ISI}$ is small (e.g. $CV_{ISI} = 0.5$; Fig. 4, upper left panel), the JNDs are very small as the population firing rate is extremely narrowly distributed, so Weber's law does not hold for independent superposition of a large population. When the population size is relatively small ($p = 100$), the Poisson process describes Weber's law the best (Fig. 4, central plot; Table 1). If the population size is very small ($p = 10$), the firing intensity should be bigger than a certain minimal value (refer to Table 1), and Weber's law holds for sub-Poisson and Poisson neural discharge processes (Fig. 4, bottom panel). Figure 5A depicts the $CV_{ISI}$ values vs different firing rates and population sizes. The value of $CV_{ISI}$ is unrealistically large (30) at a high firing rate (100 Hz) and large population size (1000). The different layers in Fig. 5A represented different values of Weber's constant $k$, and $CV_{ISI}$ increases with $k$ dramatically.

This result is interesting. For Weber's law implementation on independent population neurons, it demonstrates that a small group of neurons, rather than large-population neurons, can perform the discrimination task very well (actually $p = 10$ is sufficient for $0.5 < CV_{ISI} < 1$; Table 1). If the population size is too large ($p > 1000$), Weber's law can be satisfied only if each single neuron generates its action potential at a highly irregular process ($CV_{ISI} > 4$), which is biologically impossible. The result is consistent with the experimental observation by Britten *et al.* (1992).

However, the assumption of statistical independence among cortical neuron interactions is not realistic. Nearby cortical neurons are usually highly interconnected and share common inputs. Robust correlations among neuronal activity have been reported in a number of cortical areas from electrophysiological recordings, with an averaged correlation coefficient typically ranging from 0.1 to 0.2 (Britten *et al.*, 1992; Gawne & Richmond, 1993). What would the effect be if correlation between neurons is introduced?

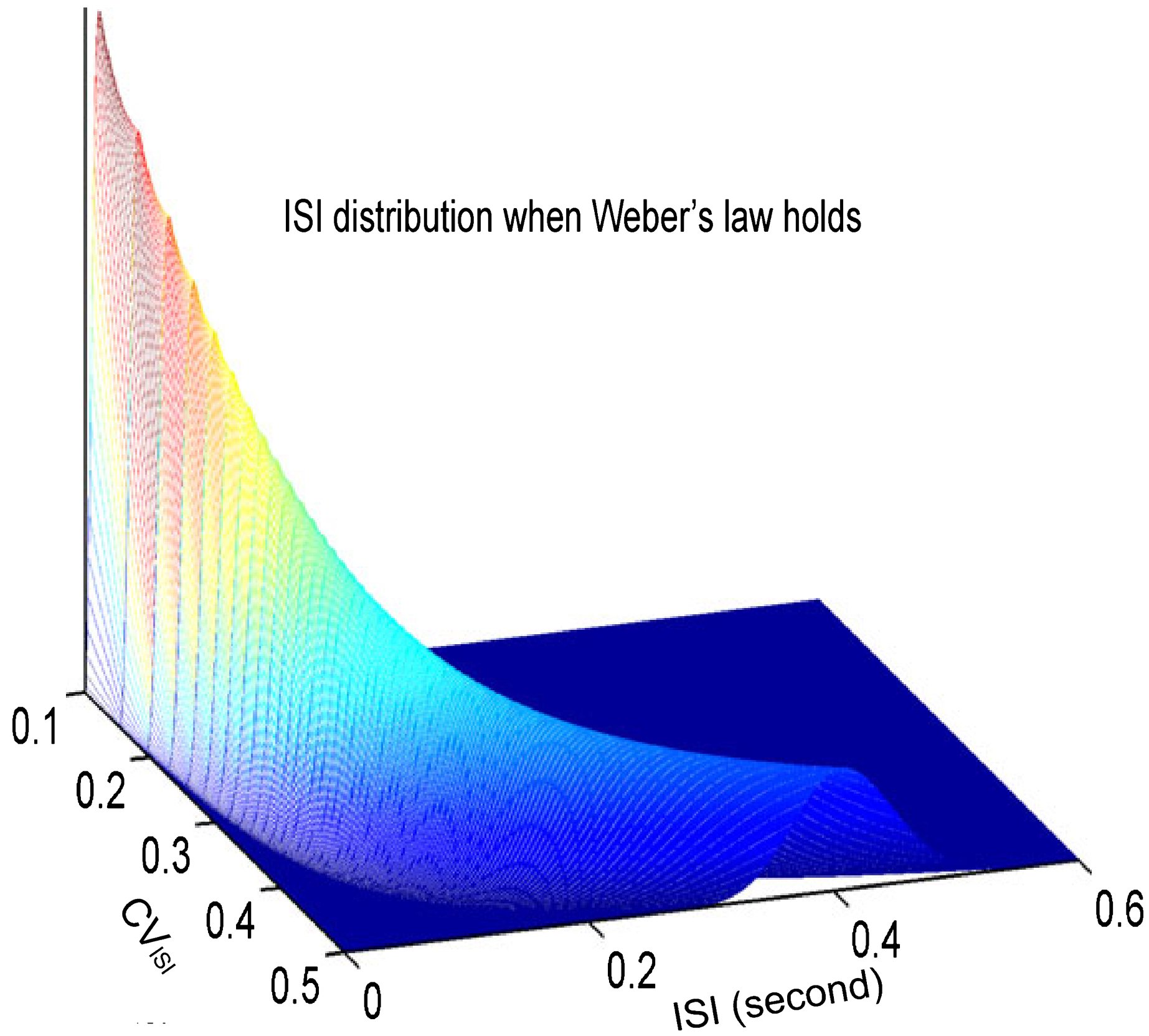

FIG. 3. Interspike interval (ISI) distribution when Weber's law is satisfied. The ISI is assumed to follow Gamma distribution, and the histogram of the ISI under Weber's equation reveals the property that the distribution is wider at low firing intensity and has big dispersion of ISI. It becomes narrower as the firing rate increases and the discharge process becomes more regular ($CV_{ISI}$ becomes smaller).

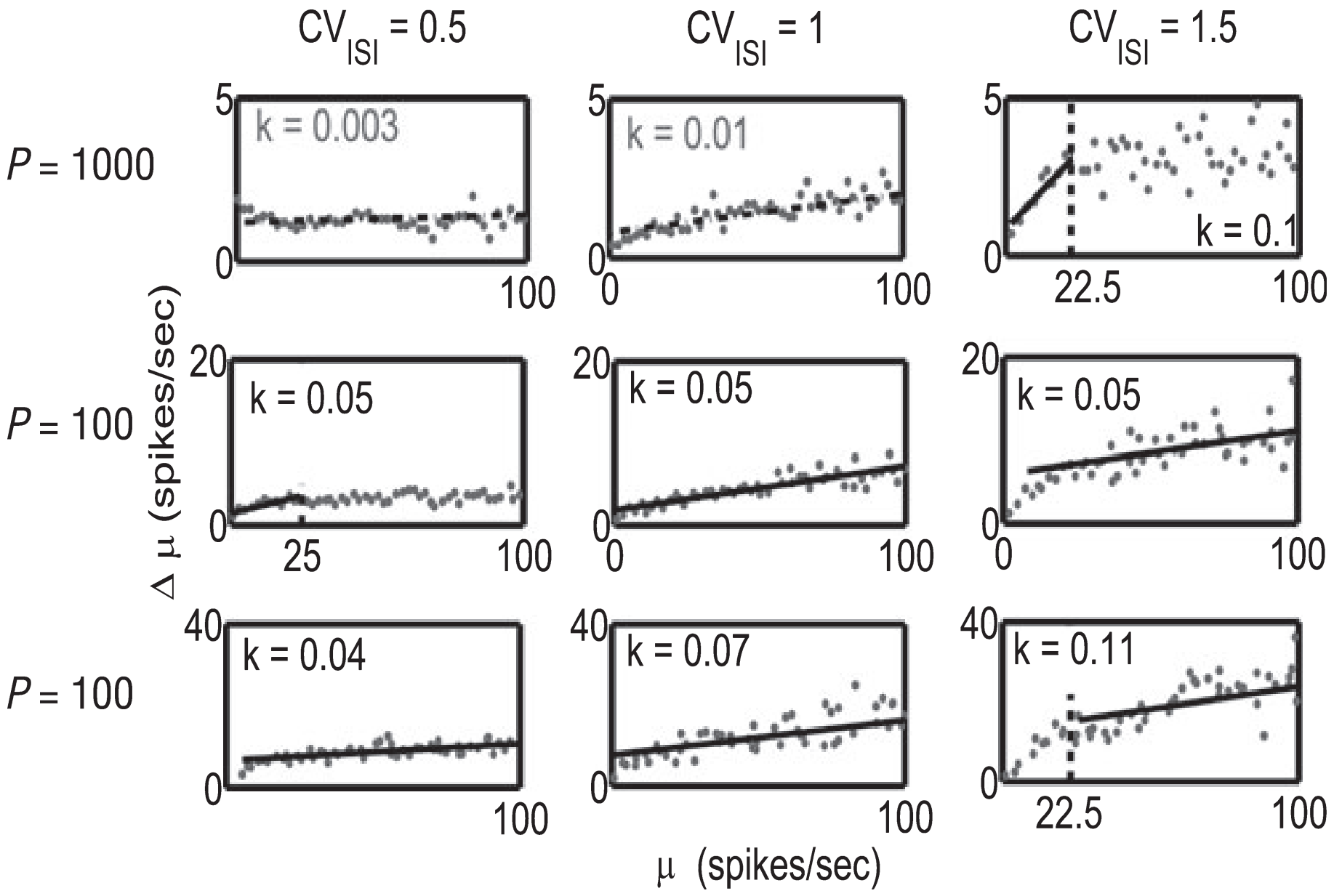

FIG. 4. The relation of JNDs and standardized reference spiking intensity $\mu$ for sub-Poisson ($CV_{ISI}$ = 0.5), Poisson ($CV_{ISI}$ = 1) and super-Poisson ($CV_{ISI}$ = 1.5) neural discharge under independent superposition process with different population size ($p$ = 1000, 100 and 10). The simulated data are fitted by linear regression and the slopes are indicated in each plot. One can see that Weber's law cannot describe the relation of JNDs and reference intensity $\mu$ well when the population size is big (upper panel) and when neuronal discharge processes are more regular (left column). The vertical dashed line represents the lower or upper bounds of the feasible range of firing intensity, and the red $k$ indicates the situation when Weber's law does not hold at all. ISI, interspike interval.

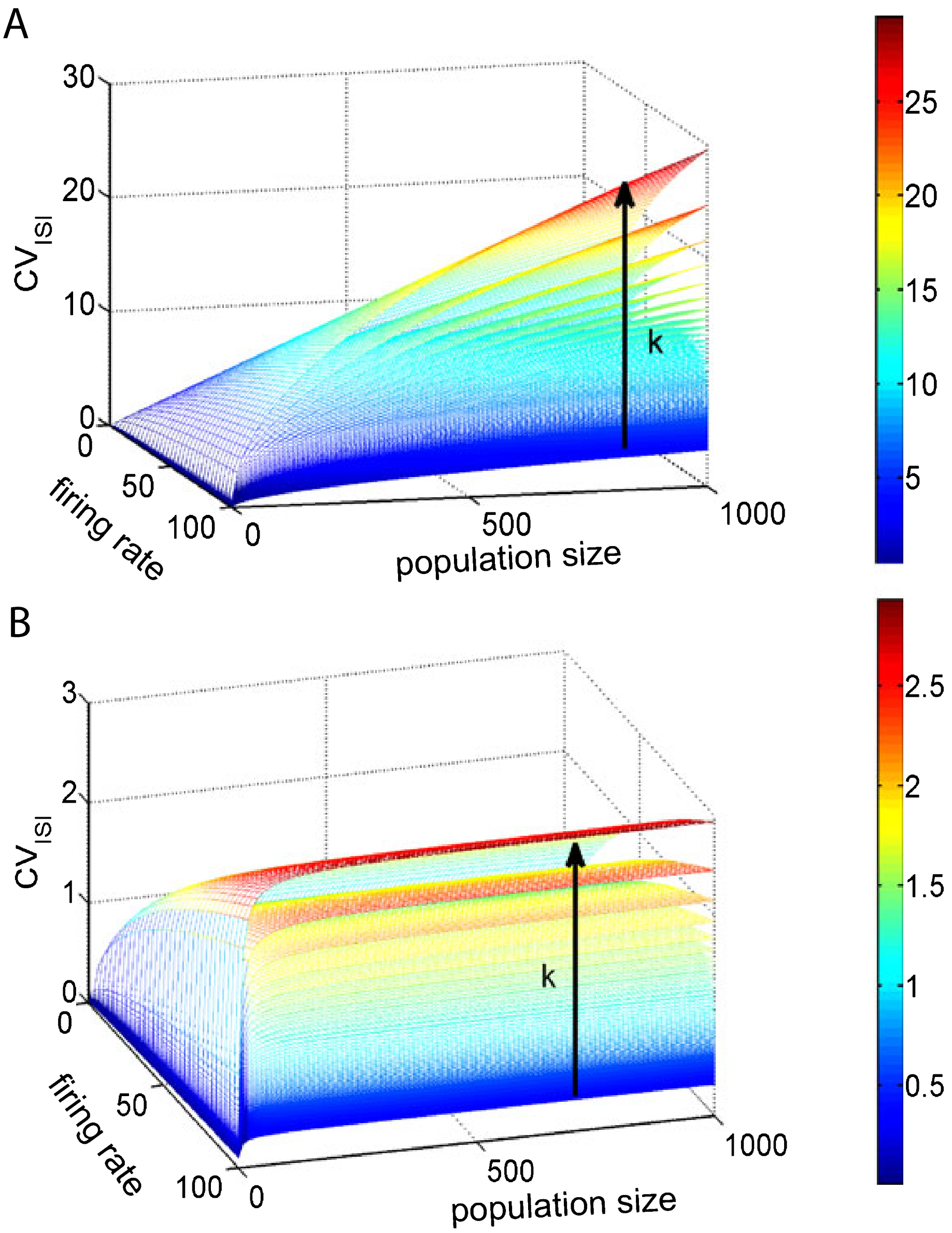

FIG. 5. The parameters relation under Weber's equation for (A) independent superposition process and (B) correlated superposition process. (A) If population neurons are independent, $CV_{ISI}$ tends to be extremely large under high firing rate and large population size. Different layers represent different values of the Weber's constant $k$. When $k$ gets bigger, $CV_{ISI}$ increases dramatically. (B) Under weakly correlated superposition process, the saturated value of $CV_{ISI}$ is about 3 when the population size tends to infinity at firing rate 100 Hz. $CV_{ISI}$ is an increasing function of the Weber's constant $k$ and the firing rate $\mu$. ISI, interspike interval.

### *Superposition of correlated neural discharge process*

Cortical cells do not generate spikes independently but, rather, the spiking activity is correlated, spatially and temporally (Smith *et al.*, 2008). Correlation arises from shared excitatory and inhibitory inputs (Morita *et al.*, 2008), either from other stimulus-driven neurons or from ongoing activities (Fiser *et al.*, 2004). In this paper, we only concentrate on the effect of spatial correlation on neural spike trains.

Assume that each neuron in the population is pair-wise correlated with the same coefficient correlation $\rho = 0.1$, and Weber's equation becomes (see Materials and methods for detailed derivation)

$$CV_{ISI} = \frac{k}{C(k+2)}\sqrt{\frac{\mu p}{(1+(p-1)\rho)}}.$$

Therefore, the corresponding Weber's range becomes [compare with Inequality (9)]:

$$\frac{\sqrt{\frac{\mu p}{(1+(p-1)\rho)}}}{100} < CV_{ISI} < \frac{\sqrt{\frac{\mu p}{(1+(p-1)\rho)}}}{10}.$$

As the population size $p \to \infty$, we have

$$\frac{\sqrt{\frac{\mu}{\rho}}}{100} < CV_{ISI} < \frac{\sqrt{\frac{\mu}{\rho}}}{10}.$$

Consequently, the values of $CV_{ISI}$ lie in between [0.45, 1] for neuronal discharge rate $10 < \mu < 200$ Hz for infinite population neurons (Table 1). The simulation results are presented in Fig. 6. When the population size $p$ is 10, the feasible range of firing rate has a lower bound – the boundary is proportional to $CV_{ISI}$ (Table 1; Fig. 6, bottom panel). When the population size $p$ increases to 100, Weber's law holds (Fig. 6, middle panel). When the population size increases from 100 to 1000, there is no significant improvement on the linear relation but, rather, the JND reveals more random property at the super-Poisson process ($CV_{ISI} = 1.5$; Fig. 6, upper-right panel). Besides, for a large population size ($p = 1000$), $\Delta\mu$ is very big ($\approx 15$ Hz) at low firing intensity ($\mu = 2$ Hz), which is not biologically plausible. The effect of the weak correlation ($\rho = 0.1$) among neurons in the superposition process can be seen from Fig. 5B. The value of $CV_{ISI}$ is bounded above by $\sqrt{\mu/10}$, which is approximately 3 for $\mu = 100$ Hz and $k = 0.3$. The different layers of the mesh represent different values of the Weber's constant $k$, and most regions of $CV_{ISI}$ lie below one for small values of $k$.

Shadlen & Newsome (1998) pointed out that the fidelity of signal transmission approaches an asymptote at 50–100 input neurons, and that there is little to be gained by involving more neurons in psychophysical performance. In our simulations, the performance of the population neurons of size 100 with correlation 0.1 matches their experimental observations very well.

### *Neural network based on competition attractor network*

In order to confirm our theoretical results ($0.5 < CV_{ISI} < 1$) derived from Weber's law, we used the model based on a competitive attractor network for decision making (Fig. 7A), firstly proposed by Brunel & Wang (2001) and subsequently examined by Deco & Rolls (2006). The $CV_{ISI}$ of neuronal ISIs for each pool of the network was calculated. The description of construction of the network is presented in the Materials and methods and supporting Appendix S1.

In this neural network model, the input spike train follows the Poisson process, and the network can make the decision following Weber's law (Fig. 7D). The measured $CV_{ISI}$ values in inhibitory, non-specific and winning pools (see Fig. 7A) are all less than one, while the losing pool has $CV_{ISI} > 1$. When the input spike train follows a different distribution (ISIs uniformly distributed between 0 and $2/F_{in}$, where $F_{in}$ is the input frequency), Weber's law no longer holds. In this case, the $CV_{ISI}$ values for the inhibitory, non-specific and winning pools become greater than one, while the losing pool has $CV_{ISI} < 1$.

The results are demonstrated in Fig. 7B–E. Figure 7B and C shows the rastergrams of randomly selected neurons from each pool in the network. Figure 7B is the rastergram when the $CV_{ISI}$ of the winning pool (pool A) is less than one. The spatio-temporal spiking activity shows the transition to the correct final state attractor. When a decision is made (after a transition period of about 700 ms), the winning pool A generates most spikes and becomes highly activated, while the losing pool B becomes relatively silent. This rastergram illustrates how pool A wins the competition and the network performs a transition to a single-state final attractor corresponding to a correct discrimination. Figure 7C is the case when the winning pool A has $CV_{ISI} > 1$. Contrary to Fig. 7B, the winning pools A with $CV_{ISI} > 1$ are very silent at the beginning, followed by a bursting for a short time (at 700 ms), then have a subsequent phase similar to Fig. 7B. The reason that causes the differences between these network behaviors can be attributed to the input distribution. Originally, the input spike train follows Poisson distribution (or equivalently, the ISI is an exponential distribution heavily distributed near zero), and this entails a very short transition time, after which the total network input is more or less constant. In the alternative case, the input spike train has its ISIs following uniform distribution, and this implies that the total input to the network is lower at the beginning of the simulation, then peaks at about time $2/F_{in}$ (which is 660 ms since $F_{in}$ is 3 Hz) and finally becomes nearly constant; this trend in the total input level is mirrored by the network firing activity. The detailed study on $CV_{ISI}$ value for each case is presented in the supporting Appendix S1.

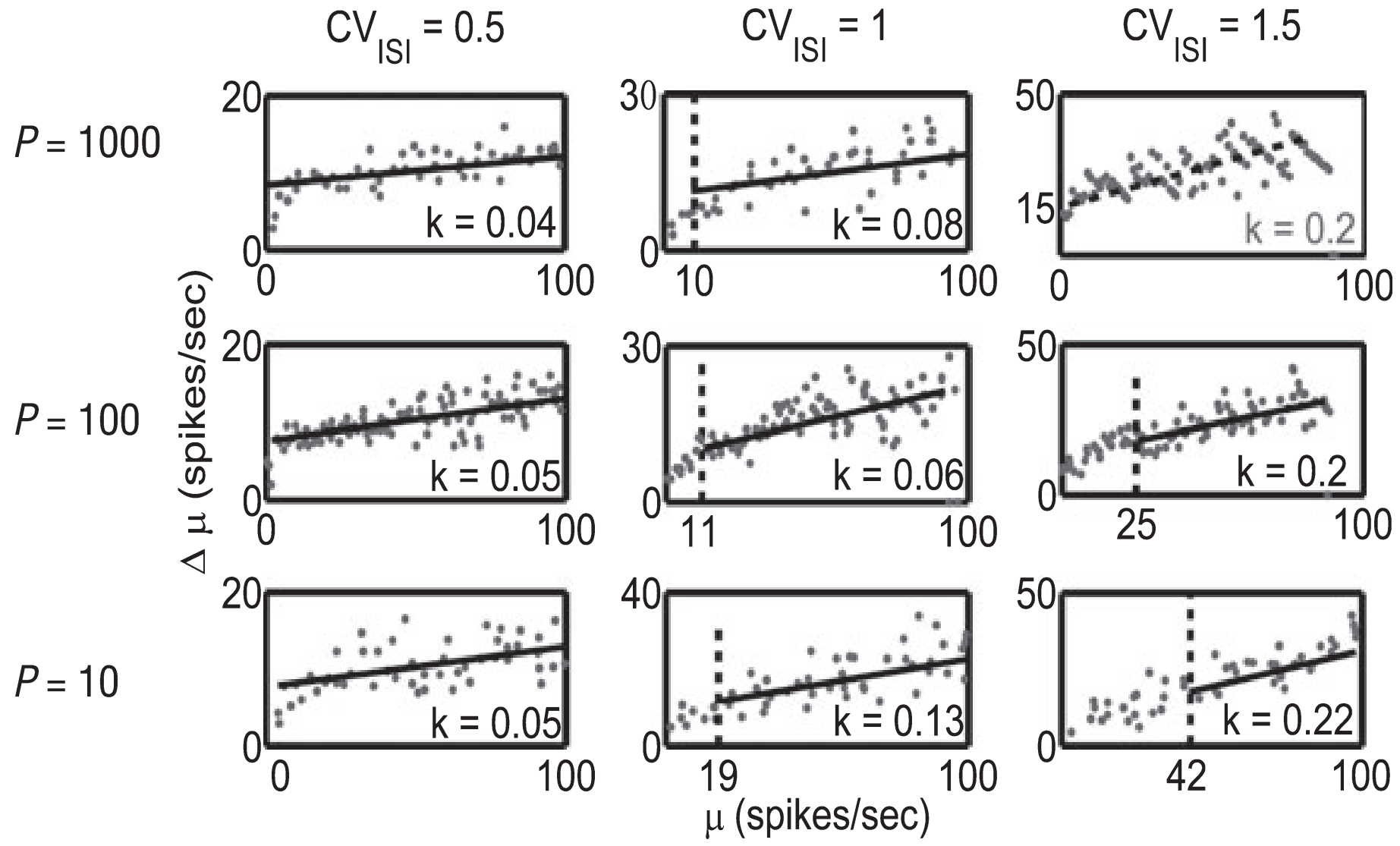

FIG. 6. The relation of JNDs and standardized reference spiking intensity $\mu$ for sub-Poisson ($CV_{ISI} = 0.5$), Poisson ($CV_{ISI} = 1$) and super-Poisson ($CV_{ISI} = 1.5$) process under correlated superposition process with different population size ($p = 1000$, 100 and 10) and correlation coefficient 0.1. The simulated data are fitted by linear regression and the slopes are indicated in each plot. The neuronal discharge rate has a lower bound for small population ($p = 10$, bottom panel), but fitting is quite good for a relatively large population size ($p = 100$ or 1000). The vertical dashed line represents the lower bounds of the feasible range of firing intensity, and the red $k$ in the top-right corner indicates the situation when Weber's law does not hold, even though the slope of the linear regression looks reasonable. Note that Weber's law holds well under small $CV_{ISI}$ (0.5), which is consistent with our theory and Table 1. ISI, interspike interval.

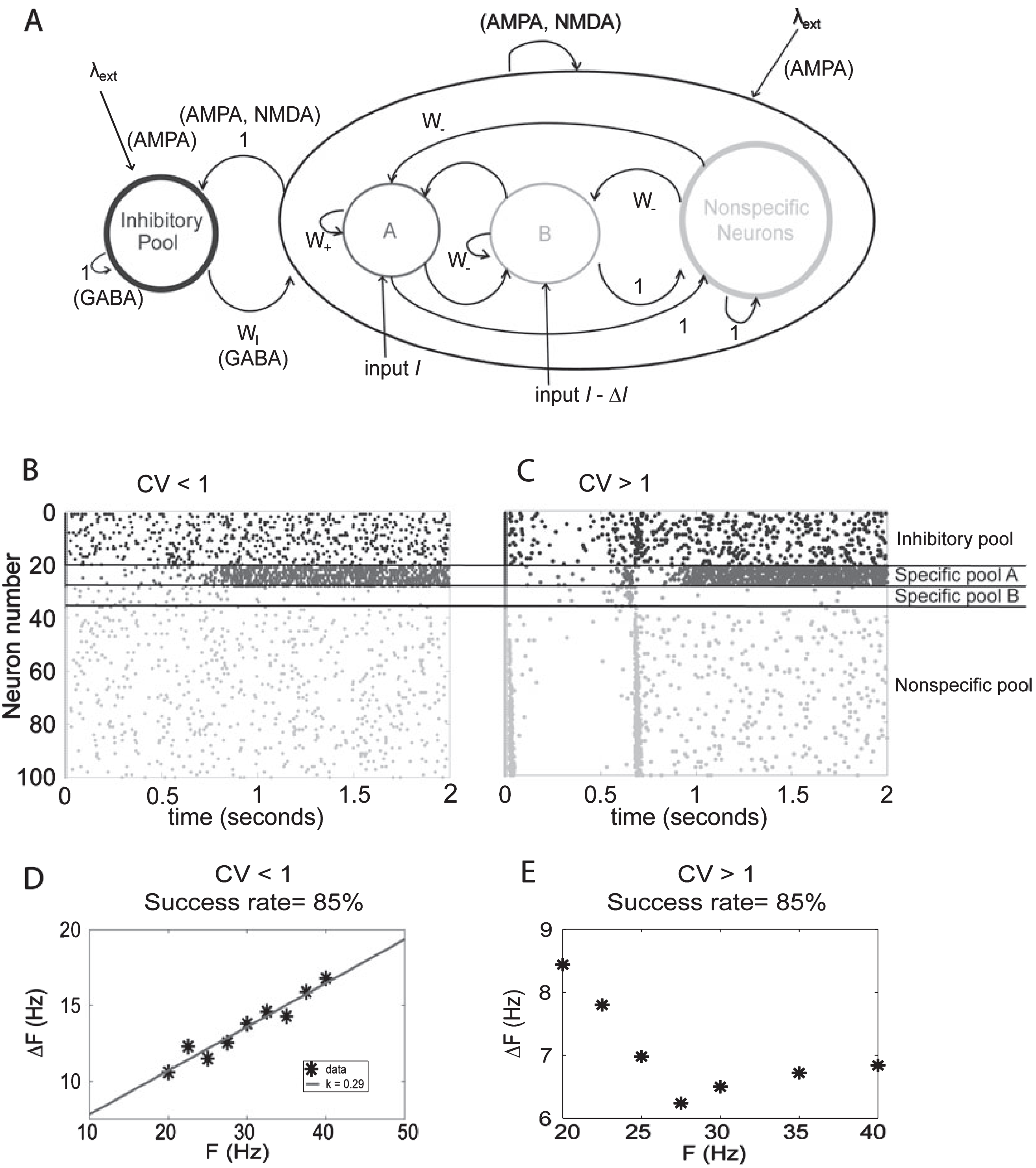

FIG. 7. The competition attractor network. (A) The architecture of the neurodynamic model for a probabilistic decision-making network. The single attractor network has two population or pools of neurons A and B. One of these pools becomes active when it wins a competition and meanwhile the other pool becomes relatively silent, which means it loses. There is also a population of non-specific excitatory neurons, and a population of inhibitory neurons. The weights connecting the different populations of neurons are shown as $w_+$, $w_-$, $w_I$ and 1. All neurons receives a small random Poisson set of input spikes $\lambda_{ext}$ from other neurons in the system. (B and C) Plots of the corresponding rastergrams of 10 randomly selected neurons for each pool in the network when (B) $CV_{ISI} < 1$ for the winning pool and (C) $CV_{ISI} > 1$ for the winning pool. Each dot corresponds to the generation of a spike, and each color represents neurons in different pools. (D and E) JND values ($\Delta I$) for the different base frequencies ($I$) when (D) $CV_{ISI}$ for the winning pool is less than one, and (E) greater than one. (D) Weber's law for the vibrotactile discrimination task. The critical discrimination Δ-value ('difference-threshold') is shown corresponding to an 85% correct performance level as a function of the base frequency *I*. The 'difference-threshold' increases linearly as a function of the base frequency. Weber's law no longer holds when $CV_{ISI} > 1$ in the winning pool, as presented in (E). AMPA, α-amino-3-hydroxy-5-methyl-4-isoxazolepropionic acid; GABA, γ-aminobutyric acid; NMDA, *N*-methyl-D-aspartate.

Figure 7D plots the linear relationship between the values of $\Delta F$ and $F$ (where $F$ is the input spike frequency) when the winning pool has $CV_{ISI} < 1$, supporting our theoretical results for Weber's law. The misclassification rate is fixed to be 15% and the linear regression has slope $k$ that varies between 0.24 and 0.32. Figure 7E is the case when the winning pool has $CV_{ISI} > 1$ and, as a result, the relationship between the reference frequencies $F$ and the values of $\Delta F$ is far from linear and is not even monotone.

## Discussion

In this paper, we derived the relationship between the mean and the standard deviation of neuronal discharge rate when Weber's law holds, and expressed the relation in terms of $CV_{ISI}$. It is found that under Weber's law ($CV_{ISI} \in [0.45, 1]$),neurons generate more regular spikes than a Poisson process. For a single neuron, the relative regular discharge process can satisfy Weber's law, but for superimposed population neurons the firing variability can be larger either with a small group of independent spiking neurons or a large group of correlated cortical cells. The findings may shed light on the theory between cortical neuronal firing property and this psychophysical law quantitatively.

### *Non-linear relation of stimulus–neuronal responses*

Cortical neurons *in vivo* usually have a non-linear response to external stimulus, and the input–output relation is commonly described by a sigmoid function (S-shaped; Fig. 8A). We carry out the analysis on this non-linear input–output relation in the supporting Appendix S1, showing that our conclusions based on the linear assumption in the Materials and methods still hold.

The relation between the $CV_{ISI}$ and neuronal firing rate $\mu$ can be obtained numerically (supporting Appendix S1, section VI) and is presented in Fig. 8 under non-linear stimulus–neuronal responses. Figure 8B is the contour plot of the $CV_{ISI}$ with respect to Weber's constant $k$ and firing rate $\mu$. The maximum value of $CV_{ISI}$ is reached at the middle value of the firing rate (about 60 Hz) because of the non-linearity of the input–output relation, and $CV_{ISI}$ is always smaller than 1 under these parameter regions. Figure 8C and D shows the illustration for the feasible range of $CV_{ISI}$ under reasonable biological parameter regions ($k \in [0.05, 0.3]$ and $\mu \in [10, 100]$).Comparing the effects on $CV_{ISI}$ ranges for non-linear input–output relation (Fig. 8C) with the linear case (Fig. 8D), the feasible range for $CV_{ISI}$ does not change much ($CV_{ISI} \in [0.17, 0.3]$for non-linear relation; $CV_{ISI} \in [0.2, 0.4]$for linear relation). Therefore, our conclusions obtained from linear stimulus–neuronal relation from the previous analysis still hold for non-linear input–output relation under Weber's law.

### *Relation between the mean and variance of neural signal*

Weber's equation [Eq.(7)], which describes the linear relation between the standard deviation and the mean of the neural firing rate, is consistent with the movement control model proposed by Harris & Wolpert (1998), who assumed that the variance of the neural signal (neuronal firing rate) is proportional to the square of the mean neural signal. There are also experimental and theoretical evidences from force production supporting the linear scaling of force signal variability (STD) with respect to the mean force level as a natural byproduct of the organization of motor neurons and muscle fibers (Jones *et al.*, 2002; Faisal *et al.*, 2008).

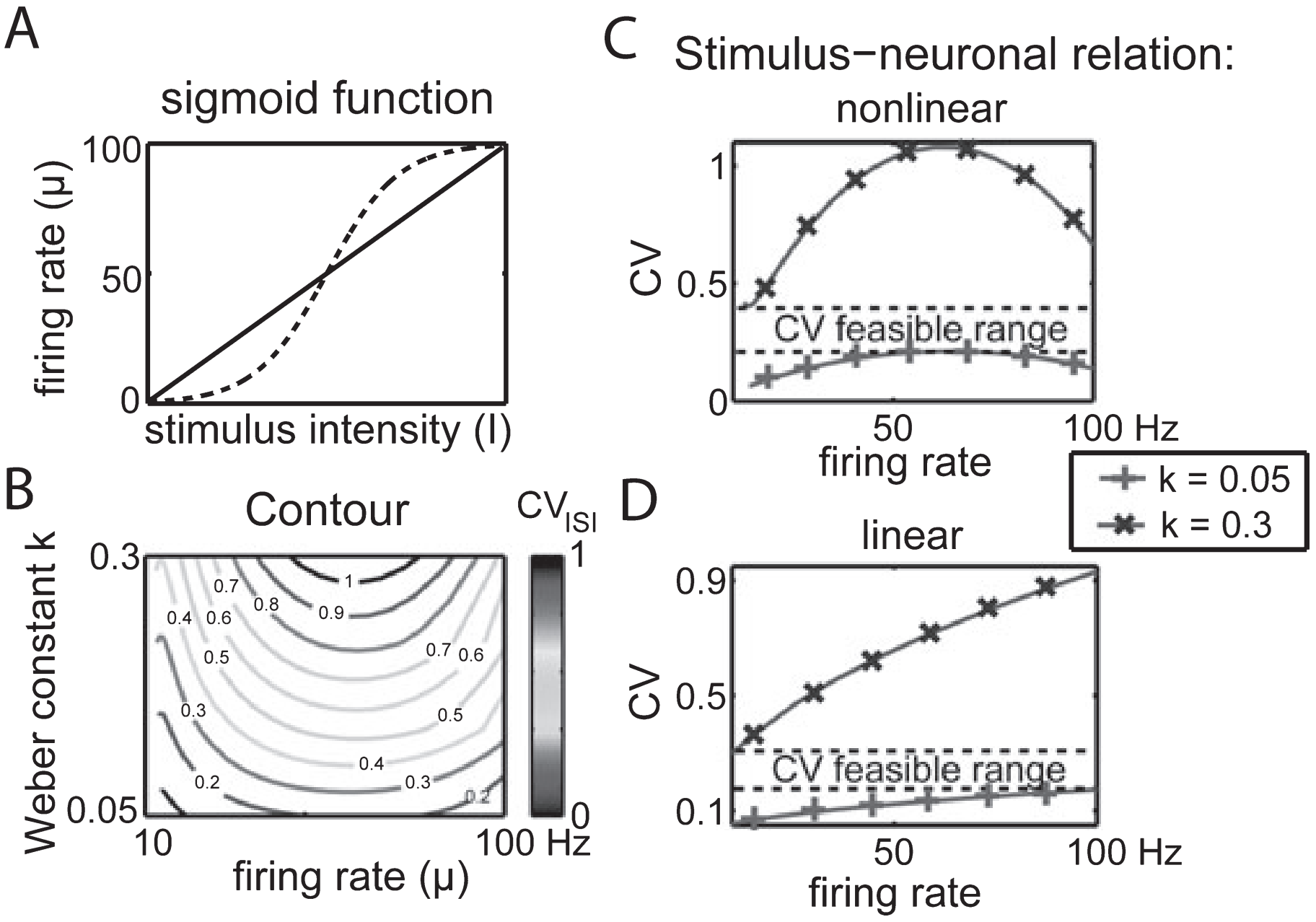

FIG. 8. Analyses on the effects of $CV_{ISI}$ range if the input–output relationship between the stimulus intensity and the neural response rate is of sigmoid function. Parameter values used for the numerical simulation are: $R_0 = 120$, $k \in [0.05, 0.3]$, $k_0 \in [0.3, 2]$. (A) Illustration for the sigmoid function compared with linear relation between the neuronal firing rate vs input stimulus intensity. (B) Contour plots for $CV_{ISI}$ values with respect to firing rate $\mu$ and Weber's constant k. (C) Numerical solution of $CV_{ISI}$ vs firing rate $\mu$ when the input–output is of non-linear relation, under different Weber's constant k (0.05 and 0.3). For every value of k and $\mu$, the feasible range of $CV_{ISI} \in [0.2, 0.4]$ is determined by the maximal and minimal value of the red line ($k = 0.05$) and blue line ($k = 0.3$). (D) When the input–output are of linear relation, the feasible range of $CV_{ISI} \in [0.17, 0.3]$, as a comparison with the non-linear case presented in (C).

Some researchers are arguing that neural discharge rate with the standard deviation ($\sigma$) linearly related to the mean ($\mu$) is a super-Poisson control signal because $\alpha = 1$ in $\sigma = \kappa\mu^{\alpha}$, while a Poisson process should satisfy $\alpha = 0.5$ (and sub-Poisson process $\alpha < 0.5$; Feng *et al.*, 2002, 2004). In fact, the definition above only considers the power $\alpha$ of the mean neural signal but neglects the scale coefficient $\kappa$. In this paper, we conclude that Weber's law implies neural discharge more regular than a Poisson process ($CV_{ISI} < 1$), even though Weber's equation matches the case $\alpha = 1$. In Weber's equation, the constant scale $\kappa$ is very small (range from 0.01 to 0.1), and it is the range of the scale $\kappa$ that determines the range of $CV_{ISI} \in [0.5, 1]$at the given power $\alpha = 1$. Hence, the scale factor $\kappa$ should also be taken into consideration to determine which process (sub-Poisson, Poisson or super-Poisson) the neuronal signal satisfies.

### *Weber's law in single neuron or system level?*

The traditional view in sensory physiology attributes to each neuron a unique role in signaling the presence of a particular feature in the visual environment (Barlow & Narasimhan, 1972). In contrast, more recent psychophysical approaches have tended to emphasize the role of large neuronal networks and pools in solving even simple perceptual problems. It has been widely accepted that subjective intensity is based on the response of population neurons, rather than a single neuron (Vega-Bermudez & Johnson, 1999). However, a most surprising finding of sensory neurophysiology is that single neurons in the visual cortex can encode near-threshold stimuli with a fidelity that approximates the psychophysical fidelity of the entire organism (Britten *et al.*, 1992; Celebrini & Newsome, 1994). This finding is understandable (in light of Weber's range for correlated superposition), which implies that psychophysical sensitivity can exceed neural sensitivity by little more, given a modest amount of correlation in the pool of sensory neurons.

In this paper, we studied both the discharge patterns of single neurons and superimposed population neurons (independent and correlated) when Weber's law is satisfied. We found that to satisfy Weber' law, a single neuron has to have a more regular discharge process while population neurons maintain large neuronal discharge variability. This interesting finding is quite consistent with the phenomenon presented in Newsome *et al.* (1989), where Newsome proposed one possible explanation that either signals from many neuronal sources are not pooled to enhance the signal strength, or the variability in the responses of similarly tuned neurons is correlated when the neuronal performance is similar or better than the psychophysical performance.

### *Variability of spatial correlation in spike train*

The correlation between neuronal spike trains depends on distance between neurons, and neuronal correlation decreases as the distance increases. Besides, correlation among ISIs can be affected by other factors as well, such as firing rate (de la Rocha *et al.*, 2007) and type of stimulus (Kohn & Smith, 2005). In this paper, we have only studied the simplest homogeneous superposition process of the population neurons by assuming that each neuron is independent (or equally correlated, neglecting the spatial effect), identical and evenly weighted. For non-homogeneous population neurons or non-stationary neuronal discharge, the situation would be much more complicated and we did not discuss it.

Moreover, the idea of linear summation of neuronal signals may not be the best way to pool neuronal activity in the cortex. A number of psychophysical studies suggest that neuronal signals contribute disproportionately to perceptual judgments. However, Britten *et al.* (1992) applied the idea of non-linearly summing the responses of members of each pool, and did not find a significant difference with linear summation among pools.

### *Argument on Weber's law*

There are many literatures studying Weber's law from different aspects on neuronal responses. Some literatures defined Weber's law as a logarithmic relationship between the strength of the stimulus and the mean response rate of the nervous system, by stating Weber's law as $\Delta\mu = \Delta I/I$, where $\mu$ denotes the mean response rate of neurons, $I$ the stimulus intensity and $\Delta$ stands for the difference. However, this logarithmic relation between the input stimulus intensity and the mean output firing rate oversimplifies Weber's law. The mean firing rate is considered to be a continuous function of stimulus intensity, and integrating on both sides of the formula yields the logarithmic relation $\mu = \log(I)$. But this continuous function is unable to perform the discrimination task in terms of probability of making the right decision. In other words, this logarithmic relation only considers the mean discharge rate but does not take into account the variability of the spike train. There is a lack of experimental evidence supporting the logarithmic relation of stimulus intensity and mean firing rate.

Even though some new psychophysical law emerged (Stevens, 1961) when the expectation of Weber's law was not fulfilled in some experimental methods, Weber's law is still widely accepted and supported by various experiments (Mahns *et al.*, 2006) in psychophysics as a basic law. One argument on this psychophysical law is the 'near-miss Weber's law', describing the observation that Weber's law holds for a majority range of stimulus intensity but fails in a certain range. Our theory can explain this observation quite well, as for a given population size ($p$) and single neuron firing property ($CV_{ISI}$), the feasible range of the firing rate (or equivalently the stimulus intensity) can be determined under Weber's range.

### *Experimental tests of Weber's law on single neuron level*

In this paper we examined how Weber's law emerged on a neuronal level, instead of on the traditional psychophysical sensitivity, through theoretical analysis and numerical simulations. Biological experiments can be designed to test Weber's law on single neuron level at various stimulus intensities using dynamic-clamp technique (Robinson & Kawai, 1993). Single neuron recordings in rat somatosensory cortex *in vitro* with tight-seal whole-cell configuration from the soma can be performed, and the cells' internal environment can be modified by directly injecting artificial synaptic step current or conductance into the cortical neurons using patch-clamp pipette. These experimental results can further testify whether our conclusions on the single neural spiking process are true if Weber's law is satisfied. This is one of our research topics to study in the near future.

## Supporting Information

Additional supporting information may be found in the online version of this article:

Appendix S1. Derivations of equations and details of the network architecture.

Please note: As a service to our authors and readers, this journal provides supporting information supplied by the authors. Such materials are peer-reviewed and may be re-organized for online delivery, but are not copy-edited or typeset by Wiley-Blackwell. Technical support issues arising from supporting information (other than missing files) should be addressed to the authors.

## Acknowledgements

We thank Dr Hugh P. C. Robinson, Prof. Edmund Rolls and Dr Wenlian Lu for their valuable comments on this paper. The work was funded by CARMEN (EPSRC, UK) and BION (EU) grant to J.F., and Warwick Postgraduate Research Fellowship and Institute of Advance Study Early Career Fellowship to J.K.

## Abbreviations

ISI, interspike interval; JND, just noticeable difference; ODE, ordinary different equation; VMR, variance-mean ratio.